\documentclass[aip,reprint,showkeys,superscriptaddress]{revtex4-2}

\usepackage{graphicx}
\usepackage{amssymb}
\usepackage{amsmath}
\usepackage{textcomp}
\usepackage{epstopdf}
\usepackage[dvipsnames]{xcolor}
\usepackage[english]{babel}
\usepackage{ulem}
\usepackage{hyperref}
\usepackage{cleveref}
\usepackage{upgreek}
\usepackage[separate-uncertainty=true]{siunitx}

\begin{document}

\title{Photorefraction Management in Lithium Niobate Waveguides: High-Temperature vs. Cryogenic Solutions}

\author{Nina~A.~Lange}
\affiliation{Institute for Photonic Quantum Systems (PhoQS), 33098 Paderborn, Germany}
\affiliation{Department of Physics, Paderborn University, 33098 Paderborn, Germany}

\author{René Pollmann}
\affiliation{Institute for Photonic Quantum Systems (PhoQS), 33098 Paderborn, Germany}
\affiliation{Department of Physics, Paderborn University, 33098 Paderborn, Germany}

\author{Michael~Rüsing}
\affiliation{Institute for Photonic Quantum Systems (PhoQS), 33098 Paderborn, Germany}
\affiliation{Department of Physics, Paderborn University, 33098 Paderborn, Germany}

\author{Michael~Stefszky}
\affiliation{Institute for Photonic Quantum Systems (PhoQS), 33098 Paderborn, Germany}
\affiliation{Department of Physics, Paderborn University, 33098 Paderborn, Germany}

\author{Maximilian~Protte}
\affiliation{Institute for Photonic Quantum Systems (PhoQS), 33098 Paderborn, Germany}
\affiliation{Department of Physics, Paderborn University, 33098 Paderborn, Germany}

\author{Raimund~Ricken}
\affiliation{Department of Physics, Paderborn University, 33098 Paderborn, Germany}

\author{Laura~Padberg}
\affiliation{Institute for Photonic Quantum Systems (PhoQS), 33098 Paderborn, Germany}
\affiliation{Department of Physics, Paderborn University, 33098 Paderborn, Germany}

\author{Christof~Eigner}
\affiliation{Institute for Photonic Quantum Systems (PhoQS), 33098 Paderborn, Germany}
\affiliation{Department of Physics, Paderborn University, 33098 Paderborn, Germany}

\author{Tim~J.~Bartley}
\affiliation{Institute for Photonic Quantum Systems (PhoQS), 33098 Paderborn, Germany}
\affiliation{Department of Physics, Paderborn University, 33098 Paderborn, Germany}

\author{Christine~Silberhorn}
\affiliation{Institute for Photonic Quantum Systems (PhoQS), 33098 Paderborn, Germany}
\affiliation{Department of Physics, Paderborn University, 33098 Paderborn, Germany}

\date{\today}
\begin{abstract}
Lithium niobate sees widespread use in nonlinear and quantum optical devices, such as for sum- and difference-frequency generation or spontaneous parametric down-conversion. In lithium niobate waveguides, nonlinear optical processes are often limited by the so-called photorefractive effect, which limits the maximum input or output powers and impacts the nonlinear spectral response. Therefore, strategies for the management of photorefractive damage are a key consideration in device design. Usually, the photorefractive damage threshold, i.e. the maximal permissible operating power, can be increased by high temperature operation of devices. This approach, however, is not applicable in cryogenic environments, which may be required for specialized applications. To better understand the impact of photorefraction in nonlinear optical applications, we study the impact of photorefraction on the phase-matching spectra of two nonlinear-optical sum-frequency generation experiments at 1) high temperatures and 2) cryogenic temperatures. Furthermore, we present an approach to reduce the impact of photorefraction which is compatible with cryogenic operation. This comprises an auxiliary light source, propagating in the same waveguide, which is used to restore phase-matching spectra impacted by photorefraction, as well as reduce pyroelectric effects. Our work provides an alternative route to photorefraction management applicable to cryogenic environments, as well as in situations with tight energy budgets like space applications.
\end{abstract}
  

\maketitle

\section{Introduction}

Lithium niobate (LiNbO$_3$, LN) has become a key material in integrated photonics, especially for electro-optic, nonlinear and quantum optical applications. Its large electro-optic and nonlinear coefficients, wide transparency range, and compatibility with mature fabrication processes make it a highly versatile platform. With the advent of thin-film lithium niobate (TFLN), the material has gained renewed attention for scalable, high-performance photonic circuits~\cite{Bazzan2015,Boes2023,Rusing2019,Qi2020,Honardoost2020,Weis1985,Zhu2021,Chen2022review}.

Nonlinear and quantum optics aim to address a wide range of challenges ranging from quantum computing, advanced sensing and spectroscopy, or secure and high-speed communication~\cite{Zhu2021,Qi2020,Roeder2024,Roeder2024b,Serino2023,Pollmann2024,Lin2025,Kaiser2019}. This often requires devices to operate reliably across widely different optical power settings, ranging from single photons to high optical powers. Furthermore, these devices must operate in vastly different environments, including room temperature laboratories, cryogenic quantum systems~\cite{bartnick2021cryogenic,Lange2022cryogenic,lange2023degenerate,Lomonte2021}, or space-based platforms~\cite{Douglass2018,Devani2020}. Each setting and application imposes strict demands on device stability, efficiency, and spectral precision.

A core challenge in nonlinear optics is that nonlinear interactions are inherently weak, meaning that strong optical fields are typically needed to drive efficient processes such as second-harmonic generation (SHG) or spontaneous parametric down-conversion (SPDC). Even in quantum optics, where single-photon-level signals are the norm, high optical powers are often involved. This is especially true in pulsed systems, quantum frequency converters, and quantum pulse gates (QPG), where strong pump fields are used to generate and manipulate weak quantum signals~\cite{Roeder2024,Roeder2024b,Serino2023,Pollmann2024,Wang2022,Wang2023,Ma2012,Ansari18,Ansari2017,Ansari2018}.

For these systems to function reliably, the devices must remain stable across all optical power levels, as well as environmental conditions. This includes maintaining consistent phase-matching conditions and spectral characteristics, which are essential for efficient and predictable operation. However, lithium niobate is known to exhibit photorefraction, which is an optically induced change in its refractive index and absorption~\cite{chen1969optically,chon1993photorefractive,pal2015photorefractive}. Photorefraction is often only considered in terms of limiting the absolute power of a device. However, especially for spectrally-engineered and quantum applications, its impacts on the spectral characteristics are equally challenging and can be observable already at $\upmu$W power levels~\cite{Ansari18}. In particular, photorefraction can both lower the device efficiency and shift the phase-matching spectrum and operating wavelength of a device, making precise wavelength targeting difficult~\cite{pal2015photorefractive}. Such accurate wavelength operation, however, is crucial for interconnects or frequency-based operations~\cite{Roeder2024,Roeder2024b,Serino2023,Pollmann2024,Wang2022,Wang2023,Ma2012,Ansari18,Ansari2017,Ansari2018}. 

The origins of photorefraction in LN are complex and involve various mechanisms, including one- and two-photon absorption on energy levels of intrinsic (e.g. Nb antisites and Li vacancies) and extrinsic (e.g. iron impurities) defects. The process is then followed by the ferroelectric photovoltaic effect, and the formation of stable charge distributions in the form of (bound) polarons~\cite{Imlau2015,Villarroel2010,Jermann1995,Furukawa2001}. Here, trapped carriers generate internal electric fields and strain that locally alter the refractive index and absorption profile of the material. The result is a dynamic and often unpredictable change in device behavior, especially under high optical intensities. Crucially, even when the absorption has not increased drastically, these refractive index profiles can alter phase-matching spectra.

In light of these challenges, many strategies have been developed to suppress and manage photorefraction in LN devices. Doping of LN, e.g. with MgO, or using \mbox{(near-)}stoichiometric LN is a common strategy to obtain a material with higher optical damage threshold~\cite{Jermann1995,Furukawa2001,Volk1994}. However, this strategy is not always applicable, for example in situations with co-doping, such as in titanium-doping for waveguides. Here, titanium occupies the same lattice sites as the Mg ions~\cite{Friedrich2017,Bocchini2025,Kollewe1992}, which can result in one dopant pushing out the other, limiting the effectiveness of this strategy. Further, while doping increases the damage threshold, photorefraction may still occur - just at increased power densities. Alternatively, operation of LN devices at temperatures in the \SI{100}{\degreeCelsius} to \SI{200}{\degreeCelsius} range is a common strategy to suppress photorefraction, as elevated temperatures increase carrier mobility and decrease the polaron lifetime. This results in a much higher optical damage threshold and allows for stable and predictable operation of devices~\cite{Imlau2015,Rams2000,Kostritskii2008}.

However, this approach is not viable in cryogenic environments, such as for integration of LN devices with superconducting nanowire detectors, or space environments~\cite{bartnick2021cryogenic,Lange2022cryogenic,lange2023degenerate,Lomonte2021,Douglass2018,Devani2020}, where energy budgets are limited and thermal control is constrained. Even for room temperature devices, for example, when integrating TFLN with silicon-based CMOS chips~\cite{Churaev2023,Lee2024,Rusing2019}, heating beyond \SI{100}{\degreeCelsius} might pose a serious limitation due to temperature constraints of the silicon CMOS electronics and photonics~\cite{Kurata2025}. As a result, there is currently no broadly effective method for suppressing photorefraction in LN, which can be applied from room temperature down to low temperatures.

We have shown before that titanium in-diffused periodically-poled waveguides in lithium niobate (Ti:PPLN) allow for the realization of cryogenic, integrated nonlinear interactions such as SHG~\cite{bartnick2021cryogenic}, SPDC~\cite{Lange2022cryogenic,lange2023degenerate,lange2025widely}, and sum-frequency generation (SFG)~\cite{lange2025widely}. However, these processes demonstrate reduced efficiencies at cryogenic temperatures, mainly limited by localized, internal electric fields disturbing the refractive index profile. These electric fields can be explained by two main causes. On the one hand, due to pyroelectricity in LN, the temperature change itself induces variations in the spontaneous polarization and thus the build-up of unscreened bound charges. On the other hand, the photorefractive threshold for tolerated optical power is significantly reduced at lower temperatures. Consequently, any in-coupled light intensity can cause additional perturbations to the refractive index distribution~\cite{lange2025widely}. Gaining insight into these aspects and providing approaches to manipulate trapped charge carriers is essential for advancing the integration of Ti:PPLN with cryogenic components. These insights can also serve as guidelines when using the TFLN plattform, where photorefraction may also play a significant role in nonlinear-optical devices~\cite{Wang2025,Xu2021}.

In this work, we investigate the impact of photorefraction in titanium-doped LN waveguides. We specifically focus on the phase-matching spectrum of SFG processes (rather than efficiency or conversion efficiency of such a process), because the phase-matching responds very sensitively to photorefractive effects, even at low powers. To investigate these effects exemplary, we use two different samples, which each represent a typical use case at high and low temperatures, respectively. The sample used at room temperature (320~K) up to 470~K is a typical QPG sample designed for pulsed operation with high peak powers alongside weak quantum signals ~\cite{Roeder2024,Roeder2024b,Serino2023,Pollmann2024,Wang2022,Wang2023,Ma2012,Ansari18,Ansari2017,Ansari2018}. In contrast, at low temperatures (7~K) we study a sample designed for relatively low power cw-driven SFG. Here, while the peak powers are much lower, an onset of photorefractive effects is already visible at relatively low powers. While the samples are not directly comparable, they show exemplary situations, where the impact of photorefraction on spectra might be a concern.

Building on this, we explore a method to suppress photorefraction under cryogenic conditions. We adapt the ``optical cleaning'' method first introduced by Kösters et al.~\cite{kosters2009optical,sturman2009optical}. They exposed the surface of a waveguide with a moving, green light beam while heating the sample to moderate temperatures. This beam enabled pushing photo-excitable electrons out of the illuminated region. While the energy of green photons is significantly less than the band gap energy, recent experiments on conductive domain walls also indicate that green light can result in increased charge carrier density or mobility~\cite{Ding2024}. This method was adapted to reduce photorefraction by Kirsch et al. \cite{Kirsch2026}, by coupling an auxiliary green laser beam together with the pump light for the nonlinear process directly into the waveguide. In this work, we demonstrate that this suppression method is both effective at and compatible with cryogenic operation temperatures.

\section{Methods and Materials}

\begin{figure*}[t]
    \includegraphics[width=0.95\linewidth]{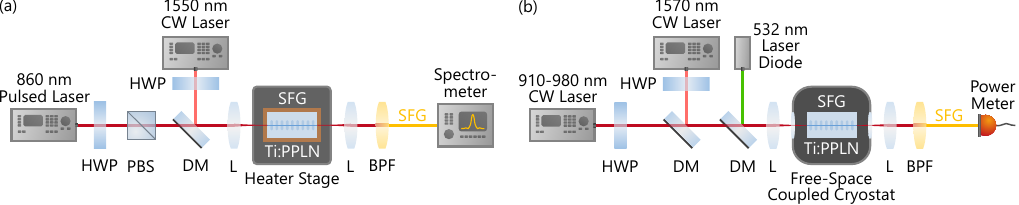}
    \caption{Experimental setup to investigate the phase-matching of a Ti:PPLN waveguide (a) at elevated temperatures, and (b) under cryogenic conditions.
    The heated waveguide is pumped with one pulsed and one continuous-wave (CW) laser to enable sum-frequency generation (SFG) and the generated beam is detected with a spectrometer.
    The cryogenic waveguide is pumped with two CW lasers to enable SFG.
    The generated beam is spectrally filtered and detected with a power meter.
    This process is combined with a CW auxiliary laser source at \SI{532}{nm} to suppress photorefractive damage.
    HWP: half-wave plate, PBS: polarizing beam splitter, DM: dichroic mirror, L: aspheric lens, BPF: band-pass filter.
    }
    \label{fig:setup}
\end{figure*}

The samples used in this study consist of titanium in-diffused periodically-poled lithium niobate (Ti:PPLN) waveguides fabricated on congruent, $z$-cut LN substrates~\cite{Sohler2008}. Here, waveguides are first defined by 80~nm Ti-strips of 5 to 7~$\upmu$m width, which are lithographically structured. The Ti-strips are then in-diffused at approximately 1060\textdegree C for 4~h. Subsequently, the waveguides are periodically-poled by electric field poling using lithographically-structured electrodes. The waveguides exhibit a mode field diameter in the range of approximately 5 to 10~$\upmu$m, depending on the exact waveguide width and operation wavelength.
Both chips contain multiple waveguides with slightly varied poling periods and waveguide widths to allow for fine-tuning of phase-matching conditions. For the high and low-temperature experiments, a different chip was used since the required poling periods for phase-matching are highly temperature-dependent.

For the high-temperature experiments we use a 7~$\upmu$m wide waveguide with a poling period of 4.35~$\upmu$m. The chip is optimized for type-II SFG between one transversal magnetically (TM) polarized field near 860~nm and a transversal electrically (TE) polarized field at 1550~nm, producing a TE polarized output around 560~nm~\cite{Serino2023,Eckstein2011,Brecht2014}. Such a sample is typically used for pulsed operation in a QPG ~\cite{Roeder2024,Roeder2024b,Serino2023,Pollmann2024,Wang2022,Wang2023,Ma2012,Ansari18,Ansari2017,Ansari2018}.
In contrast, the waveguide used in the low-temperature experiments has a width of \SI{5}{\um} and features a larger poling period of \SI{9.90}{\um}. This poling is designed for a type-0 SFG process involving two pump wavelengths near \SI{1570}{nm} and \SI{960}{nm}, producing a signal around \SI{595}{nm}, whereby all fields are TM polarized. More details on the cryogenic SFG process can be found in Ref.~\cite{lange2025widely}. As discussed below, using a purely TM-based type-0 process at low temperatures makes the sample particularly sensitive to charge build-ups at surfaces~\cite{lange2025widely}, allowing to study the impact of the auxiliary laser. While the nonlinear optical (NLO) processes are not exactly identical, they represent typical use cases for NLO processes at each temperature. 

The high-temperature experiments are conducted using a tunable mode locked Ti:Sapphire laser (Coherent Chameleon Ultra)  emitting 150~fs pulses around 860~nm as the TM pump field. 
This laser is mixed with $\approx1$~mW of the TE input fixed at 1550~nm from a tunable CW  laser (EXFO T100S-HP), imitating the operating conditions of a QPG (see Fig.~\ref{fig:setup}~(a)) \cite{Serino2023,Eckstein2011,Brecht2014}.
Since the TM pump is at $\approx3$~THz much broader than the phase-matching function, the whole output spectrum can be recorded at once with a grating spectrograph (Andor Shamrock SR-500i) equipped with an EMCCD camera (Andor Newton 970P).

The chip is mounted on a custom-built temperature-controlled heater stage driven by a PID controller (Oxford Mercury).
Usually, this stage allows for tuning of the phase-matching condition via thermal control.
More importantly in our case, this setup enables the study of photorefraction suppression as a function of temperature.
Spectral measurements are taken at various combinations of heater temperature and pump laser power to map the influence of thermal and optical conditions on photorefraction behavior.

The low-temperature experiments follow a similar configuration but are adapted for cryogenic operation (see Fig.~\ref{fig:setup}~(b)).
In this setup, a tunable CW pump laser (EXFO T500S) is set to a fixed wavelength of 1570~nm and combined with a second tunable CW laser around 960~nm (TOPTICA Photonics CTL 950). Since both lasers are CW, their bandwidths are much narrower than the phase-matching spectrum. For this reason, the complete spectrum cannot be measured at once. Instead, the second laser is scanned while the output signal around 595~nm, generated via sum-frequency generation, is monitored with a power meter.
Here, the chip is mounted in an optical cryostat (Attocube attoDRY800) with accessible optical windows for in- and out-coupling.
The cryostat allows cooling of the sample down to approximately 7~K.
The cryostat is further modified to include a heater which is mounted close to the sample. This heater regulates the speed of the cooldown, so that the cooling rate does not exceed \SI{1}{K/min}. 

To investigate photorefraction suppression at low temperatures, an auxiliary laser at 532~nm with an optical power of \SI{1}{mW} is introduced. This wavelength is selected based on the prior work by Kösters et al.~\cite{kosters2009optical,sturman2009optical}, and Kirsch et al. \cite{Kirsch2026}, who demonstrated the reduction of photorefraction at room temperature in LN, when applying green light.

\section{High-Temperature Suppression of Photorefraction}

Photorefraction in lithium niobate is an optically induced change of the local refractive index.
In integrated nonlinear devices, even small changes in the refractive index will modify the phase-matching function, i.e. the spectral response of the device.
Larger changes in the refractive index can even change the spatial mode profile and lead to coupling between formally orthogonal spatial modes, potentially resulting in the complete breakdown of device functionality.
\begin{figure*}[t]
    \includegraphics[width=1\textwidth]{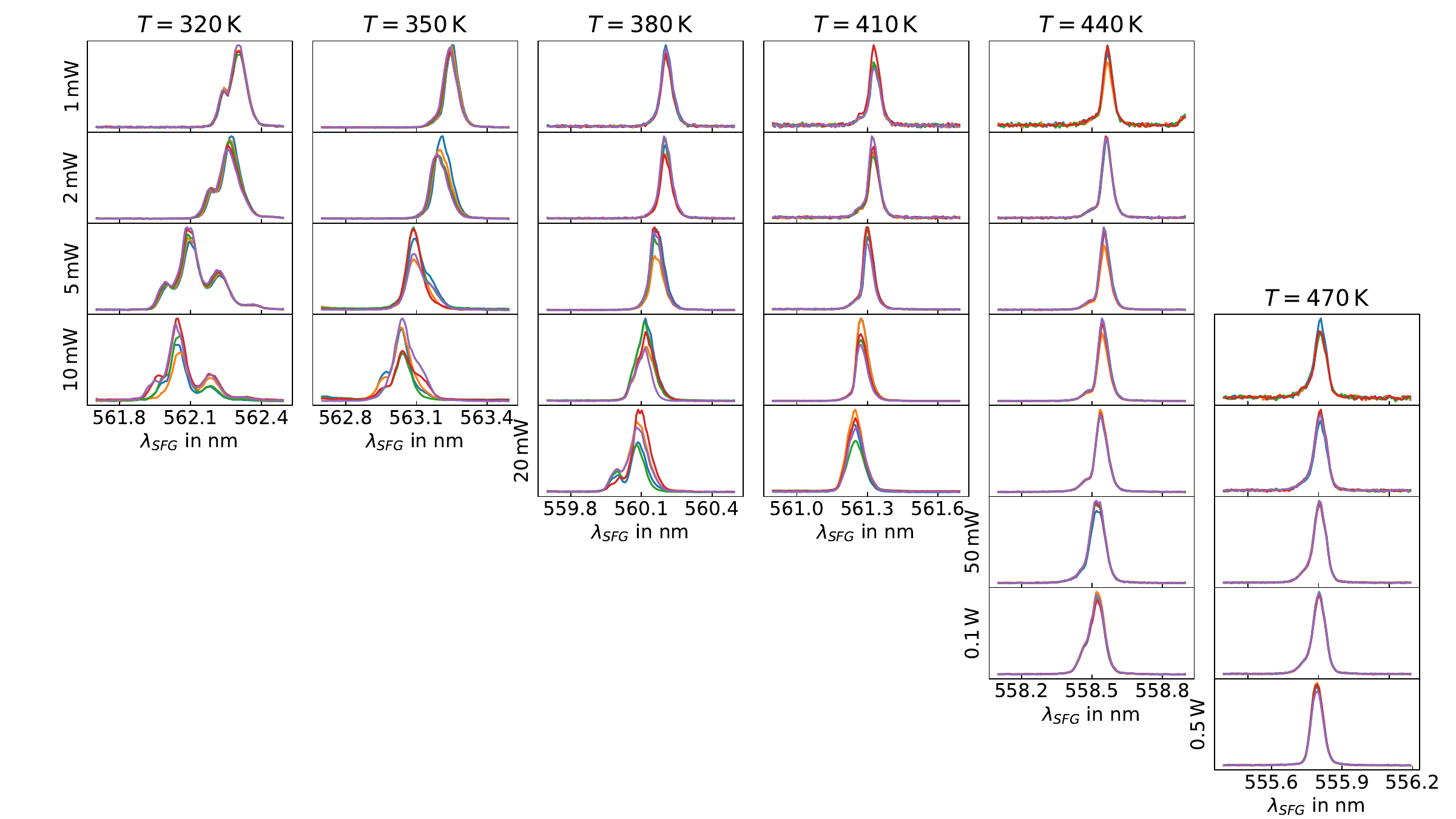}
    \caption{Repeated measurement of the phase-matched SFG spectrum of the high temperature waveguide.
    At low power (first row), the waveguide shows an unperturbed spectral shape and only minuscule changes between repeated measurements (different colored lines).
    At low temperature (left column), as the power is increased, the spectrum shifts towards shorter wavelengths, gets perturbed and shifts between measurements.
    With increased temperature, the onset of both effects shifts to higher pump powers.
    }
    \label{fig:ht_spectra}
\end{figure*}

In the high temperature experiments we slowly heat the sample and keep it at the desired temperature for at least 4~h to ensure any pyroelectric charge is dissipated and the optomechanical setup is in thermal equilibrium.
Due to the shift in the phase-matching with temperature, we adjust the TE pump wavelength in the near infrared (NIR) for each setting to ensure optimal overlap.
We keep the wavelength and input power of the telecom field constant at $1550$~nm and $\approx1$~mW, respectively, while varying the NIR power from $<1$~mW at low temperatures to $>800$~mW at the highest temperature using a motorized wave plate and a polarizing beam splitter.
The NIR power is continuously monitored via a reflection from a dichroic mirror.
At each power setting we take five spectra of the produced light in the span of 10~s to record fluctuations on short timescales.
Fig.~\ref{fig:ht_spectra} shows these spectra for selected NIR powers.
Due to the vastly different input powers and variations in coupling efficiency, we separately normalized each set of spectra.
This normalization allows us to focus on the changes in the spectral shape of the phase-matching independent of conversion efficiency.

\begin{figure*}[t]
    \includegraphics[width=1\textwidth]{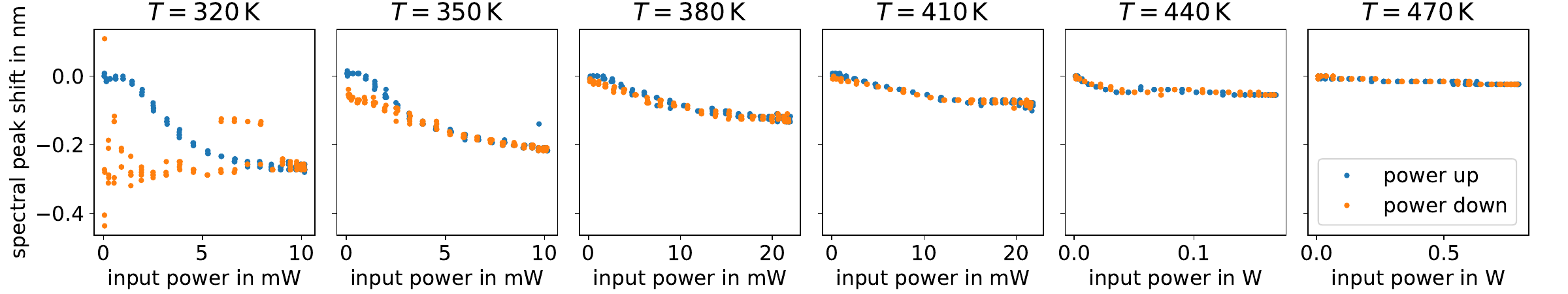}
    \caption{
        Relative position of the SFG maximum over the pump power for the high temperature waveguide.
        With increasing pump power (blue), the center of the phase-matched SFG shifts towards shorter wavelengths.
        As the temperature is increased, this shift decreases in magnitude.
        At temperatures over 380~K the phase-matching returns to its original position within the runtime of the measurement as the pump power is decreased (orange).
        At the lower temperatures the spectrum stays disturbed, showing significant hysteresis.
    }
    \label{fig:ht_shift}
\end{figure*}

To mitigate the risk of permanent optical damage to structures with more narrow band phase-matching on the same chip, we deliberately choose a structure with a phase-matching broadened by fabrication imperfections.
At all temperatures we observe a shift of the phase-matched SFG to shorter wavelengths with increasing pump power, which is reduced significantly at high temperatures (see Fig.~\ref{fig:ht_shift}). This shift to shorter wavelength is typical for photorefractive effects~\cite{Wang2025}.
We took data both for increasing and decreasing pump powers, shown in blue and orange, respectively.
At temperatures above 380~K both traces are virtually identical, as any photorefractive modulation is sufficiently suppressed before the next spectrum is recorded.
At lower temperatures, however, a significant hysteresis is observed indicating the long stability of generated charges.
Especially at 320~K, the phase-matching spectrum did not recover within the 4.5~min long measurement.

At temperatures below 380~K we coupled sufficient power to also observe significant broadening, as well as fluctuations of the SFG spectrum.
Whereas at 410~K we only observe intensity fluctuations which are then completely suppressed above 440~K.
The prominent second peak observable at 320~K for input powers above 5~mW is most likely caused by the NIR field occupying the first order spatial mode TM$_{10}$. 
As expected, this breakdown of the SFG process is shifted to larger input powers for increased temperature.

\section{Managing Photorefraction at Cryogenic Temperatures}
 
Based on our demonstration experiment at high temperatures, one could see that even at moderate powers, onsets of photorefraction can already impact the phase-matching spectra. 
The cryogenic operation affects the waveguiding and nonlinear conversion efficiencies mainly due to an interplay between the pyroelectric effect and photorefraction.
The pyroelectric properties of lithium niobate cause a temperature dependence of the inherent spontaneous polarization. The cooldown to cryogenic temperatures thus results in a change in the polarization strength which manifests itself in the form of unscreened bound-charges. For $z$-cut lithium niobate, the strong polarization causes charge carriers to accumulate on the top and bottom surface of the waveguide chip. This results in localized electric fields which are aligned in the $z$-direction, i.e. parallel to the crystal axis. Since the in-diffused waveguides are close to the top surface, the electric fields disturb the refractive index profile due to the electro-optic effect. These electric fields mainly affect the TM-polarized beam which is polarized parallel to the crystal axis~\cite{chen1969optically}. For this reason, we characterize a cryogenic type-0 NLO process which only employs TM polarization since it provides specific insights into the influence of photorefraction and pyroelectricity.
Under ambient conditions, the electric fields can often be compensated by surrounding atmospheric charge carriers, which cannot happen in the vacuum chamber of a cryostat. This means that arising electric fields can only vanish through discharges within the material. The low charge carrier mobility~\cite{Zahn2024} at cryogenic temperatures makes discharges unlikely. Previous research has shown that the number of spontaneous discharge occurrences is significantly reduced for temperatures below \SI{100}{K}~\cite{thiele2024pyroelectric}. Consequently, any non-discharged charge carriers are ``frozen in'' and will disturb the refractive indices at cryogenic temperatures. For this reason, we perform the cooldown slowly, limiting the cooling rate to \SI{1}{K/min}, so that accumulated charges have more time to discharge before their mobility vanishes at cryogenic conditions. However, charges will remain.

Due to the reduced charge carrier mobility, photorefraction in lithium niobate becomes especially prominent in cryogenic environments. The high-temperature suppression of photorefraction has shown that the threshold level for accepted optical power can be raised substantially when increasing the operation temperature as shown in the previous chapter. This implies that the threshold will be significantly decreased at cryogenic temperatures. Moreover, the strongly reduced charge carrier mobility makes it much more challenging to counteract induced damage.
This ``freeze in'' of de-localized charges remains unchanged as long as no external energy is supplied. This can either happen thermally by heating the waveguide, or optically by the additional coupling of a high-energy laser beam.

\begin{figure*}[t]
    \includegraphics[width=1\linewidth]{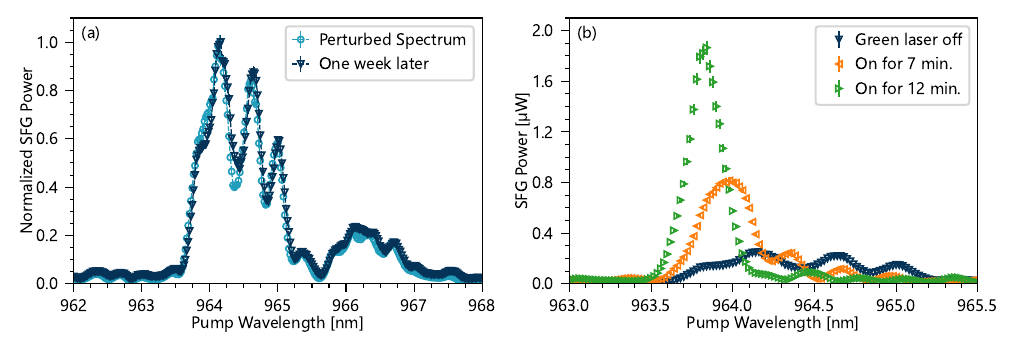}
    \caption{Phase-matched SFG spectrum of the cryogenic waveguide. (a) The waveguide shows a perturbed spectrum which differs from the optimal shape due to the build-up of electrical charges in the waveguide region. This spectral shape does not change when the waveguide is kept cold and non-illuminated for one week, indicating that the charges are not moving under cryogenic conditions. (b) As soon as the light from a green laser diode is simultaneously coupled to the waveguide, the spectrum starts to change. The dark blue data points correspond to the same measurement as shown in (a), when the green light is still turned off. The spectral shape approaches the ideal $\mathrm{sinc}^2$ shape and the SFG power increases when illuminating the waveguide for 12 minutes with the green light.}
    \label{fig:LT_PerturbedSpectrum}
\end{figure*}
All cryogenic SFG experiments are performed in the same waveguide and at \SI{7}{K}. In the first experiment, we set the input power of the telecom laser to \SI{24.4(8)}{mW}, measured in front of the in-coupling lens. The pump power of the second laser was set comparatively high to \SI{23.6(8)}{mW}. The cryogenically measured phase-matching spectrum exhibits an initially distorted shape (see Fig.~\ref{fig:LT_PerturbedSpectrum}~(a)). The spectral shape strongly differs from a $\mathrm{sinc}^2$ profile, which is expected for an ideal nonlinear process, as for example seen in Fig.~\ref{fig:ht_spectra}. We attribute this distortion to the combined effects of pyroelectric charge build-up during the cooldown and photorefractive damage due to excessive optical pump power. To test the temporal stability of the disturbed phase-matching spectrum, the waveguide was subsequently kept at a constant temperature with no in-coupled light for one week. We then repeated the measurement, but this time using lower input power values. The power of the telecom laser was kept constant from now on at \SI{20.0(6)}{mW}. We expect photorefractive damage to mainly result from the higher energy pump laser and the generated SFG signal in the visible range. For this reason, we reduced the power of the second laser strongly to \SI{1.80(6)}{mW} to avoid additional photorefractive damage. The spectrum exhibits no significant changes compared to the measurement one week prior. This suggests that, in the absence of sufficient thermal or optical energy, the dislocated charges remain immobilized under cryogenic conditions for at least one week.

To explore the possibility of actively mobilizing the ``frozen in'' charges, in a second experiment we kept the pump power at the low level and additionally coupled the light from the green laser diode into the waveguide. We measured the phase-matching spectrum repeatedly and show the induced changes after \SI{7}{\minute} and \SI{12}{\minute} in Fig.~\ref{fig:LT_PerturbedSpectrum}~(b). Remarkably, the spectral shape began to evolve within a few minutes, after it remained unchanged for one week when no light was coupled. After only \SI{7}{\minute} of auxiliary laser exposure, the spectrum approached the expected $\mathrm{sinc}^2$ shape with less pronounced side-lobes. We further saw a blue shift in the wavelength that yielded maximum SFG power.
Following \SI{12}{\minute} of illumination, the maximum generated power increased significantly from \SI{0.25(1)}{\uW} to \SI{1.87(6)}{\uW} and the phase-matching presents a shape that is close to the ideal $\mathrm{sinc}^2$ shape. These findings indicate that sufficient optical energy can mobilize the trapped charge carriers, even under cryogenic conditions. As a consequence, they can recombine or migrate into new positions and thus enable modification of the phase-matching characteristics. Interestingly, this increase in mobility can already be achieved with sub-band-gap illumination, which was also observed with photoconductivity experiments on conductive domain walls~\cite{Ding2024}. In this regard, the green laser wavelength fits well with the known energy level of the bi-polaron, which can be excited with green laser light and broken into the more mobile free and bound polaron in LN~\cite{Reichenbach2018,Imlau2015}.

\begin{figure*}[t]
    \includegraphics[width=1\linewidth]{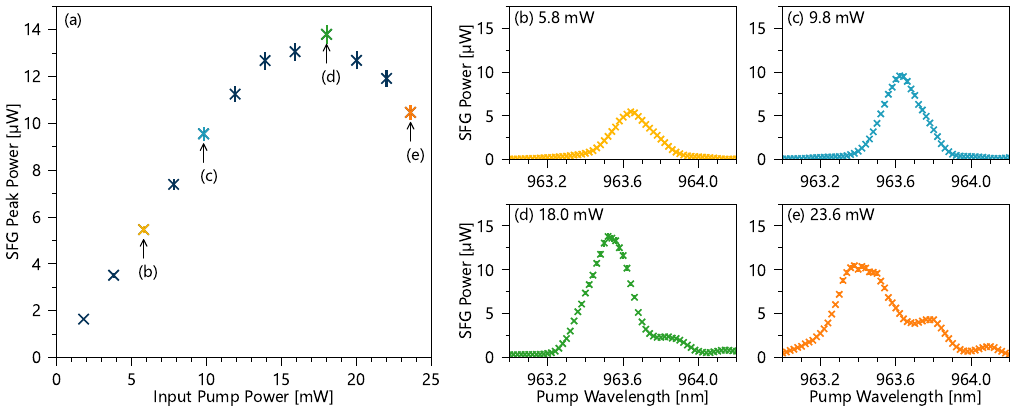}
    \caption{Increasing the pump power of the SFG process results in photorefractive damage to the cryogenic waveguide. Any initial damage was compensated before with the light from a green laser diode (see Fig.~\ref{fig:LT_PerturbedSpectrum}). (a) The green light is turned off and the pump power is gradually increased. The peak power of the SFG process first follows the trend of the increasing pump power and then drops again. (b) - (e) Individual spectra measured for different pump powers. Increasing the power results in broadening of the spectrum and the evolution of side peaks.}
    \label{fig:LT_PumpPowerIncrease}
\end{figure*}
Next, we switched off the green laser and investigated the effect of incrementally increasing the pump power on the phase-matching. Fig.~\ref{fig:LT_PumpPowerIncrease}~(a) displays the dependence of the SFG peak power on the pump power of the tunable laser. The SFG peak power initially increases linearly, as expected for the low pump power regime without inducing photorefractive effects. However, when the pump power exceeds approximately \SI{15}{mW}, the slope reaches a maximum before the generated power starts to decrease.
This effect can be explained by photorefraction rather then pump-depletion or saturation, because the generated power (\SI{12}{\uW}) is much lower then the provided power at either the telecom (\SI{20}{mW}) and pump wavelength (\SI{15}{mW}). This assumption is supported by the observation of increasing distortions in the phase-matching spectrum similar to the high-temperature observations. While the individual spectra at lower pump power (see Fig.~\ref{fig:LT_PumpPowerIncrease}~(b)~-~(c)) exhibit a single, well-defined peak at the same central wavelength, spectra recorded at higher pump power (see Fig.~\ref{fig:LT_PumpPowerIncrease}~(d)~-~(e)) show a shift towards shorter wavelengths and the emergence of side peaks. The amount of distortion clearly grows as the pump power is increased. These results indicate that the cryogenic waveguide exhibits an evident threshold for optical power. This threshold was observed for the combination of an input pump power of \SI{18}{mW} at approx. \SI{960}{nm} and a generated SFG power of about \SI{14}{\uW} around \SI{595}{nm}. Exceeding this threshold causes serious impairment of device performance.

\begin{figure*}[t]
    \includegraphics[width=1\linewidth]{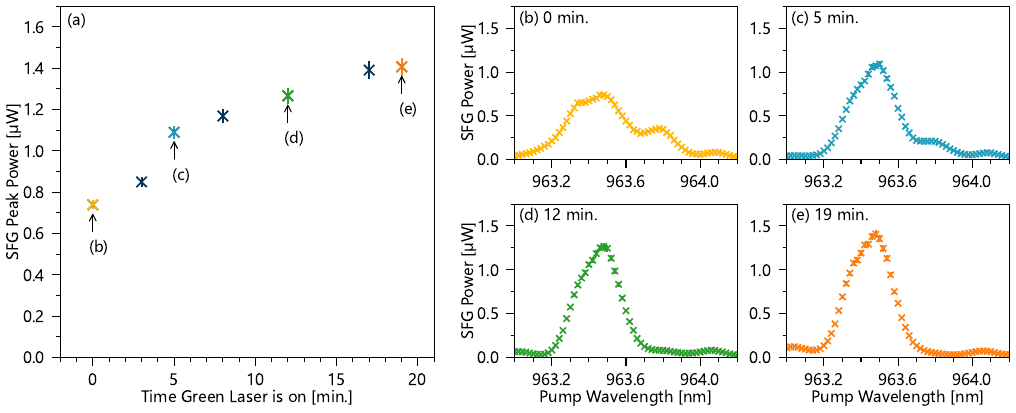}
    \caption{The photorefractive damage, which was induced by high optical pump power (see Fig.~\ref{fig:LT_PumpPowerIncrease}), can be partly soothed by repeating the simultaneous coupling of the green laser diode. (a) Keeping the green light turned on for about 19 minutes results in an increase of the SFG peak power. (b) SFG spectrum measured before the green light was turned on. (c) - (e) Individual spectra measured over time after turning on the green light. The spectral shape improves when coupling the green laser for 19 minutes.}
    \label{fig:LT_HealingGreen}
\end{figure*}
After photorefractive damage was induced using excessive pump power levels, we show that this damage can be partly mitigated again by repeating the simultaneous coupling of the green laser. For this, the input power of the tunable laser was again reduced to \SI{1.8}{mW}, before the green laser was switched on and the phase-matching spectrum was monitored over time. The temporal changes in SFG peak power are shown in Fig.~\ref{fig:LT_HealingGreen}~(a). We observe an increase in SFG peak power and a simultaneous disappearance of the previously formed side peaks, which can be seen from the individual spectra in Fig.~\ref{fig:LT_HealingGreen}~(b)~-~(e). The variations seem to slowly saturate after about \SI{19}{\minute}. However, in comparison to the first application of green light, the phase-matching spectrum is not fully recovered. The SFG peak power does not reach the same level as before and this time, the central peak shifts to even shorter wavelengths. This partial recovery suggests that while green illumination can relax a significant portion of the photorefractive damage, some residual effects persist after repeated cycles of damage and healing.

Moreover, we want to point out that a consistent blue shift was observed throughout the measurements. It has been shown before that photorefractive damage can result in a blue shift of the phase-matched wavelength~\cite{pal2015photorefractive}. However, we observed the blue shift not only for the increasing pump power experiment (which was meant to induce photorefraction), but also for the first green light application which was meant to relax the damage. Specifically, the maximum SFG power initially occurred at \SI{964.16(2)}{nm}, shifted to \SI{963.84(2)}{nm} after first green exposure, further to \SI{963.38(2)}{nm} after increasing the power, and finally remained almost unchanged at \SI{963.48(2)}{nm} during the repeated green coupling. These observations indicate that the photorefractive effects in the cryogenic waveguide are only partially reversible. Although the coupling of the green laser increases the generated power and improves the spectral shape, the blue shift implies that the charge carrier distribution and refractive index changes do not fully recover to its initial state.

\section{Conclusion}
While waveguides in lithium niobate represent a well-established technology for integrated quantum optical devices, the photorefractive effect constitutes a limitation for the tolerated optical power. Intrinsic material defects can absorb part of the optical power, which results in redistribution of the charge carriers. The arising localized electric fields cause disturbances of the refractive index profile, thus decreasing nonlinear conversion efficiencies and inducing changes in the phase-matching spectrum.

We demonstrated the effect of photorefraction when performing sum-frequency generation in two different waveguides. The first waveguide was actively heated above room temperature, and the second waveguide was operated cryogenically at about \SI{7}{K}. The first experiment clearly showed that high optical power levels induce significant changes in the phase-matching spectra. We have shown that this damage can be prevented by operating the waveguide at high temperatures due to increased charge carrier mobility. Obviously, heating of the waveguide is not compatible with cryogenic operation. However, the cryogenic investigation showed that ``frozen-in'' charges can be actively mobilized by additionally coupling a green laser to the waveguide. Our results indicate that the photorefractive damage can be partly relaxed, but some residual effects remain. The observed blue shift suggests that the damage is not fully reversible at cryogenic temperatures when coupling the currently used green laser diode.

We suggest that future work could explore the optimization of the presented suppression protocol. This could include the modification of the auxiliary laser beam, such as using pulsed illumination, different wavelengths, and fine-tuning the required optical power and exposure duration. Moreover, a thorough theoretical study of the cryogenic material parameters and the exact energy level of the bi-polaron can support optimization of the auxiliary laser properties. Relaxation of the charge carriers may be further improved by the controlled application of external electric fields or the use of conductive coatings on the waveguide surface. Combining different approaches can help to develop a routine in-situ ``healing procedure'' to maintain the device performance over a longer period of time. Managing photorefraction in cryogenic lithium niobate waveguides would clearly advance the quantum photonic application possibilities for this widely used material platform.

\begin{acknowledgments}
We acknowledge financial support from the Deutsche Forschungsgemeinschaft (Grant No. 231447078-TRR 142). Portions of this manuscript were drafted with the assistance of OpenAI's ChatGPT (model GPT-4). The final text was reviewed and edited by the authors.

\end{acknowledgments}


\section*{Data availability}
Data underlying the results presented in this paper are not publicly available at this time but may be obtained from the authors upon reasonable request.

\bibliography{paper}

\end{document}